\documentclass[aps,pra,twocolumn,showpacs,floatfix]{revtex4}
\usepackage{epsfig}
\usepackage{graphicx}
\usepackage{dcolumn}
\usepackage{amsthm,amsmath}

\begin{document}
\title{Multipolar Black Body Radiation Shifts for the Single Ion Clocks}
\author{$^1$Bindiya Arora, $^2$D. K. Nandy and $^2$B. K. Sahoo \footnote{Email: bijaya@prl.res.in}}
\affiliation{$^1$IISER Mohali, Sector 81 Mohali, P.O. Manouli, 140306, India \\
$^2$Theoretical Physics Division, Physical Research Laboratory, Ahmedabad-380009, India}
\date{Received date; Accepted date}
 
\begin{abstract}
Appraising the projected $10^{-18}$ fractional uncertainty in the optical
frequency standards using singly ionized ions, we estimate the black-body
radiation (BBR) shifts due to the magnetic dipole (M1) and electric quadrupole
(E2) multipoles of the magnetic and electric fields, respectively. Multipolar
scalar polarizabilities are determined for the singly ionized calcium (Ca$^+$) 
and strontium (Sr$^+$) ions using the relativistic coupled-cluster method;
though the theory can be exercised for any single ion clock proposal. The expected
energy shifts for the respective clock transitions are estimated to be 
$4.38(3) \times 10^{-4}$ Hz for Ca$^+$ and $9.50(7) \times 10^{-5}$ Hz for
Sr$^+$. These shifts are large enough and may be
prerequisite for the frequency standards to achieve the
foreseen $10^{-18}$ precision goal.
\end{abstract}

\pacs{31.25.-v, 32.10.Dk, 06.30.Ft, 32.80.-t}
\maketitle

\section{Introduction}

Optical transitions with ultranarrow frequencies in the single positively charged
ions which use advanced laser cooling and trapping techniques are of current 
interest for frequency standards \cite{gill,diddams,rosenband}. Some 
of the successful single ion optical atomic clocks are Hg$^+$ 
\cite{oskay,bize}, Ca$^+$ \cite{matsubara}, Sr$^+$ \cite{margolis, dube},
Al$^+$ \cite{chou1,rosenband}, Yb$^+$ \cite{schneider} etc., among these 
the fractional uncertainty in Hg$^+$ and Al$^+$ have already reached 10$^{-17}$
\cite{rosenband}. A new range of experiments are also proposed for
other ions like Ba$^+$ \cite{sherman,sahoo1}, Ra$^+$ \cite{sahoo2},
In$^+$ \cite{becker,wang}, Yb$^+$ \cite{hosaka} etc. More precise 
frequency standards will open-up
possibilities to study the underlying physics related to fundamental
constants, probing new elementary physics, more importantly it can
help in improving the present day global positioning systems and also
tracking of deep-space probes \cite{gill,bize,chou2,dzuba,flambaum}.
 
One of the major fortifications to attain smaller fractional 
uncertainties in the optical frequency standard measurements using ions
is the accurate measurement of black-body radiation (BBR) shifts. The 
considered standard frequency is shifted from the atomic resonance value
due to the interaction of the ion with the external stray electromagnetic fields
present in and around the experimental apparatus \cite{mizushima}. The BBR shift 
is caused by the isotropic field radiated due to finite 
temperature of the apparatus \cite{autler,farley,mizushima}.

The dominant contribution to the BBR-induced energy shift is from the 
electric dipole (E1) component of the radiation field; which have been 
gauged by many groups for a number of ions using the relativistic theories
\cite{BBR-arora-ca,BBR-arora-sr,BBR_Porsev,sahoo2,sahoo3,sahoo4,kallay}.
However, there is absolutely no rigorous estimate of the BBR shift due 
to higher multipoles for any proposed scheme. Following derivations
for E1 BBR shift in \cite{autler,farley,mizushima}, Porsev and Derevianko
have given a generalized derivation \cite{BBR_Porsev} to deduce the BBR
shifts spawned by any multipole component of the electromagnetic field.

Two of the proficiency gadgets for the optical frequency standards are with 
a single calcium ion ($^{43}$Ca$^+$) \cite{champ}  trapped in a Paul trap 
and with a strontium ion ($^{88}$Sr$^+$) confined in an endcap trap
\cite{Madej,margolis}. The considered clock transitions in these ions are the
$s_{1/2} \rightarrow d_{5/2}$ transitions operating in the optical regime, the
principles are similar to the proposed Ba$^+$ \cite{sherman,sahoo1}, 
Ra$^+$ \cite{sahoo2} and Yb$^+$ \cite{schneider, hosaka} based frequency standards. A major
advantage of using $^{43}$Ca$^+$ ion is that the radiation required for
cooling, repumping, and clock transition is easily produced by non-bulky solid
state or diode laser ~\cite{champ}. The reported frequency measurements of the
transition for frequency standard in $^{88}$Sr$^+$ have achieved a spectral
resolution of better than 1.5 Hz~\cite{Madej,margolis,margolis03,Madej99}.   
As has been proclaimed, these experiments have the dexterity to diminish
the relative systematic uncertainties to a level of $10^{-17}$ or below
\cite{sahoo2,margolis}. In such a scenario, it is compelling to estimate
the BBR shifts caused by the higher multipoles, especially through the M1 and 
E2 channels for the experiments comprising the $s_{1/2} \rightarrow d_{5/2}$ 
transitions as in the above ions. Such an attempt was made by Porsev and 
Derevianko \cite{BBR_Porsev} for divalent atoms like Mg, Ca, Sr, and Yb.
In this paper, we extend the work to  preview  the BBR shifts
commenced by the M1 and E2 multipoles in the ions; particularly for the
$^{43}$Ca$^+$ and $^{88}$Sr$^+$ ions where the efficacious experiments are in 
progress to reduce the uncertainties over their previous measurements 
\cite{matsubara,champ,Madej,margolis,margolis03,Madej99}.

The paper is organized as follows: First, we discuss the inception of the BBR
shifts which embodies the M1 and E2 contributions of the radiation field. Then,
we discuss the method of calculation in the $^{43}$Ca$^+$ and $^{88}$Sr$^+$ 
ions subsequently presenting the results for the above multipolar contributions
to the BBR shifts in these ions before recapitulating the work. Unless
stated otherwise, we use the conventional system of atomic units, a.u.,
in which $e, m_{\rm e}$, $4\pi \epsilon_0$ and the reduced Planck
constant $\hbar$ have the numerical value one.

\section{Theory}
The interaction Hamiltonian between an electron in an atomic system 
with the external propagating electromagnetic field in the Coulomb gauge coupling is given by
\begin{eqnarray}
V(r, \omega) &=& - c \boldmath{\alpha} \cdot \text{\bf A}(r, \omega) \nonumber \\
             &=& - c (\boldmath{\alpha} \cdot \boldmath{\hat \epsilon}) \exp{(i {\bf k} \cdot \text{\bf r})},
\label{eqn1}
\end{eqnarray}
where $\boldmath{\alpha}$ is the Dirac matrix in operator form, $\omega$ 
is the angular frequency of the field and ${\bf k} = k \hat k$ and $\boldmath{\hat 
\epsilon}$ are its wave vector and polarization direction, respectively.
In the presence of this interaction, the energy shift that can occur 
for an atomic energy level $\vert \Psi_n \rangle$ with energy $E_n=\omega_n$
can be approximated to \cite{autler, farley}
\begin{eqnarray}
\delta E_n(\omega ) = \frac{1}{2} \sum_{m \ne n} |V(r, \omega)|^2 \left (
\frac{\omega_n - \omega_m}{(\omega_n - \omega_m)^2 -\omega^2} \right ) .
\label{eqn2}
\end{eqnarray}
Using
the multipolar expansion and in terms of the traditional multipole
moments $Q_{LM}^{\lambda}$ , we have \cite{johnson}
\begin{eqnarray}
(\boldmath{\alpha} \cdot \boldmath{\hat \epsilon}) \exp{(i {\bf k} \cdot {\bf r})} &=& 
- \sum_{L M} \frac{ k^L i^{L+1+\lambda}}{(2L+1)!!} [\text{\bf Y}_{LM}^{\lambda}(\hat{k}) \cdot \boldmath{\hat \epsilon} ] \nonumber \\
&& \sqrt{\frac{4 \pi (2L+1)(L+1)}{L}} Q_{LM}^{\lambda},
\nonumber \\
\label{eqn3}
\end{eqnarray}
where $\lambda=1$ and $\lambda=0$ represents the electric and magnetic
multipoles, respectively.

Using the relations ${\bf B} = \nabla \times {\bf A}$ and ${\bf E} = \frac{i 
\omega}{c} {\bf A}$ and combining all the above expressions, the energy
shift for an isotropic field after averaging over $\omega$ for all the
polarization and propagation directions is given as~\cite{farley,BBR_Porsev}
\begin{eqnarray}
\delta E_n^{(\lambda, L)} &=& -\frac{(\alpha T)^{2L+1}}{2J_n+1}\sum_m|<\Psi_n||Q_L^{(\lambda)}||\Psi_m>|^2 \nonumber \\ && \times \ F_L\left(\frac{\omega_{mn}}{T}\right),
\label{eqn7}
\end{eqnarray}
where $J_n$ is the angular momentum of state $\vert \Psi_n \rangle$ and
\begin{eqnarray}
F_L(y)&=& \frac{1}{\pi}\frac{L+1}{L(2L+1)!!(2L-1)!!} \nonumber\\
&\times &\int_0^{\infty}\left( \frac{1}{y+x}+\frac{1}{y-x}\right)\frac{x^{2L+1}}{e^x-1} dx.
\end{eqnarray}
Here argument $y=\omega_{mn} /T=(\omega_n - \omega_m)/T$. The function $F_L(y)$ is a 
universal function applicable to all the atoms with argument $y$ depending on 
the range of the atomic parameters. These functions were first introduced by 
Farley and Wing~\cite{farley} in the E1 case and are extended in a general form
in this work to emanate simpler forms for the energy shift expressions.
In the limit $|y|>>1$ which corresponds to the case when the transition 
energy is much larger than the temperature ($T$) is of our current interest.

Substituting values from Eq. (\ref{eqnf}) in Eq.(\ref{eqn7}), the BBR shift for $L=1$ can be expressed as
\begin{equation}
\delta E_n^{(\lambda, 1)} = - \frac{1}{2} \left (\frac{8\pi^3\alpha^3 (k_BT)^4}{45(2J_n+1)}\right ) \sum_m\frac{|<\Psi_n||Q_1^{(\lambda)}||\Psi_m>|^2}{\omega_{mn}}. 
\end{equation}

Now at the room temperature (300 K), the BBR shift for the E1 channel can be given as
\begin{eqnarray}
\delta E_n^{E1} &=& -\frac{1}{2} \frac{4\pi^3\alpha^3}{15}(k_BT)^4 \alpha_n^{E1} \nonumber \\ 
    &=& -\frac{1}{2} \langle E_{E1}^2( \omega ) \rangle \alpha_n^{E1} \nonumber \\
    &=& -\frac{1}{2} (831.9 \rm{V/m})^2\left[\frac{\rm{T(K)}}{300}\right]^4 \alpha_n^{E1},
\end{eqnarray}
whereas for the M1 channel can be reduced to
\begin{eqnarray}
\delta E_n^{M1} &=& -\frac{1}{2} \frac{4\pi^3\alpha^5}{15}(k_BT)^4 \alpha_n^{M1} \nonumber \\
    &=& -\frac{1}{2} \langle B_{M1}^2( \omega ) \rangle \alpha_n^{M1} \nonumber \\
&=& -\frac{1}{2}(2.77 \times 10^{-6} \rm{Tesla})^2\left[\frac{\rm{T(K)}}{300}\right]^4 \alpha_n^{M1}, 
\end{eqnarray}
where $\alpha_n^{E1}$, $\alpha_n^{M1}$, $\langle E_{E1}^2( \omega ) \rangle$ and 
$\langle B_{M1}^2( \omega ) \rangle $ are the scalar E1 polarizability, scalar
M1 polarizability, the averaged E1 induced electric field and the averaged M1
induced magnetic fields, respectively.

Similarly, substituting values from Eq. (\ref{eqn16}) in Eq. (\ref{eqn7}), 
the BBR shifts for $L=2$ comes out as
\begin{equation}
\delta E_n^{(\lambda, 2)} = - \frac{1}{2}\left ( \frac{16 (\alpha \pi)^5 (k_BT)^6 }{945(2J_n+1)}\right ) \sum_m\frac{|<\Psi_n||Q_2^{(\lambda)}||\Psi_m>|^2}{\omega_{mn}},
\end{equation}
which corresponds to the E2 and M2 channels.

Therefore, the BBR shift at the room temperature for the E2 channel is given by
\begin{eqnarray}
\delta E_n^{E2} &=& -\frac{1}{2} \frac{8 (\alpha \pi)^5(k_BT)^6 }{189(2J_n+1)} \alpha_n^{E2} \nonumber \\
    &=& -\frac{1}{2} \langle E_{E2}^2( \omega ) \rangle \alpha_n^{E2} \nonumber \\
    &=& -\frac{1}{2} (7.2 \times 10^{-3} \rm{V/m})^2\left[\frac{\rm{T(K)}}{300}\right]^6 \alpha_n^{E2},
\end{eqnarray}
where $\alpha_n^{E2}$ is the scalar E2 polarizability and  $\langle E_{E2}^2( \omega ) \rangle $ is the averaged E2 induced electric field.

Accumulating all the above expressions, the BBR shifts due to both the M1 and 
E2 channels for an atomic transition $\vert \Psi_f \rangle \rightarrow \vert 
\Psi_i \rangle$ at room temperature can now be given in the forms
\begin{eqnarray}
\delta E_{f \rightarrow i}^{M1} &=& -\frac{1}{2}(2.77 \times 10^{-6} \rm{Tesla})^2\left[\frac{\rm{T(K)}}{300}\right]^4  \nonumber \\ &&\times \left (\alpha_f^{M1} - \alpha_i^{M1}\right ),
\label{eqn19}
\end{eqnarray}
and
\begin{eqnarray}
\delta E_{f \rightarrow i}^{E2} &=& -\frac{1}{2} (7.2 \times 10^{-3} \rm{V/m})^2\left[\frac{\rm{T(K)}}{300}\right]^6  \nonumber \\ && \times \left (\alpha_f^{E2} - \alpha_i^{E2}\right ),
\label{eqn20}
\end{eqnarray}
respectively.

\section{Method of calculation}

In order to determine M1 and E2 polarizabilities in the considered systems, 
which have closed-shell configurations with one valence electron each, 
we first calculate the Dirac-Fock (DF) wave function ($|\Phi_0\rangle$)
for the corresponding closed-shell configurations, then append the valence
orbital ($n$) to define a new reference state (i.e. $\vert \Phi_n \rangle = 
a_n^{\dagger}|\Phi_0\rangle$) in the Fock space.  
To obtain the exact atomic wave functions (ASFs), the core, core-valence and
valence correlation effects are accounted by defining wave operators
$\Omega_c$, $\Omega_{cn}$ and $\Omega_n$, respectively. i.e. the
exact ASF $|\Psi_n \rangle$ with a valence orbital $n$  is given by
\begin{eqnarray}
|\Psi_n \rangle &=& a_n^{\dagger} \Omega_c |\Phi_0 \rangle + \Omega_{cn} |\Phi_n\rangle + \Omega_n |\Phi_n \rangle.
\label{eqn21}
\end{eqnarray}
We adopt the relativistic coupled-cluster (RCC) method in the Fock-space
to determine ASFs. With $T$ and $S_n$ representing the core electrons and
 core electrons with the valence electron excitation operators, respectively,
the above ASF in the RCC framework can be encapsulated in a form 
\cite{sahoo1,sahoo2,sahoo3,sahoo4, mukherjee}
\begin{eqnarray}
|\Psi_n \rangle &=& e^T \{1+S_n\} |\Phi_n \rangle ,
\label{eqn22}
\end{eqnarray}
where the ASF for the closed-shell configuration is given by $\Omega_c$
defining $|\Psi_0\rangle=\Omega_c|\Phi_0\rangle = e^T |\Phi_0 \rangle$,
$\Omega_{cn} = e^{T(n)} |\Phi_0 \rangle$ and $\Omega_n=e^T S_n|\Phi_n\rangle$. 
We have mentioned $T(n)$ for $\Omega_{cn}$ to emphasize on the fact that the
$T$ operator in this case excites at least one of the core electrons to the
valence orbital. All these correlation effects are coupled in the course of 
the ASF determination.

The equations determining the coupled-cluster amplitudes and energies are
accustomed in compact forms as
\begin{eqnarray}
\langle \Phi_0^* |\{\widehat{He^T}\}|\Phi_0 \rangle &=& \delta_{0,*} \Delta E_{corr}
\label{eqn23}
\end{eqnarray}
and
\begin{eqnarray}
\langle \Phi_n^*|\{\widehat{He^T}\} \{1+S_n\}|\Phi_n\rangle &=& \langle \Phi_n^*|1+S_n|\Phi_n\rangle \nonumber \\ && \langle \Phi_n|\{\widehat{He^T}\} \{1+S_n\} |\Phi_n\rangle \nonumber \\
 &=& \langle \Phi_n^*|\delta_{n,*}+S_n|\Phi_n\rangle \Delta E_n, \ \ \ \ \ \ \ \
\label{eqn24}
\end{eqnarray}
where the notation $(* =1,2)$ represents for the excited hole-particle
states, $\widehat{He^T}$ denotes the connected terms of the Dirac-Coulomb
(DC) Hamiltonian ($H$) with the $T$ operators, $\Delta E_{corr}$ and
$\Delta E_v$ are the correlation energy and attachment energy (also equivalent
to negative of the ionization potential (IP)) of the electron of orbital $v$,
respectively. It can be noticed that the reference states $|\Phi_0 \rangle$
in Eq. (\ref{eqn23}) and $|\Phi_n \rangle$ in Eq. (\ref{eqn24})
contain different number of particles, hence the Hamiltonian used in the
respective equations describe different number of particles in our Fock
space representation. We have considered contributions only from the singly
and doubly excited states along with some important valence triple excitations which
are included perturbatively; the approach is known as CCSD(T) method in the
literature (e.g. see \cite{sahoo1,sahoo2,sahoo3,sahoo4,mukherjee}).

To calculate the scalar polarizabilities, it is precedence to adopt
an approach similar to \cite{sahoo3, sahoo4} or it would be prudent
to follow a procedure given in \cite{kallay}. Such approaches may be required
for achieving better accuracies. Accumulating from the previous expressions,
the general definition of the scalar polarizability is given by
\begin{eqnarray}
\alpha_n^{Q_{\lambda}^L} &=& \frac{1}{\alpha^{2(\lambda-1)}}\frac{2}{(2L+1)}\frac{1}{2J_n+1}\nonumber \\
&& \times \sum_m\frac{\left|\left<\Psi_n||Q_{\lambda}^L||\Psi_m\right>\right|^2}{E_n - E_m} .
\label{eq25}
\end{eqnarray}

Procuring the derivations in Eq. (\ref{mateq}), we can now write
\begin{eqnarray}
\alpha_n^{Q_{\lambda}^L}  &=& \alpha_n^{Q_{\lambda}^L}(c) + \alpha_n^{Q_{\lambda}^L}(nc) + \alpha_n^{Q_{\lambda}^L}(n),
\label{eq26}
\end{eqnarray}
where $\alpha_n^{Q_{\lambda}^L}(c)$, $\alpha_n^{Q_{\lambda}^L}(nc)$ and 
$\alpha_n^{Q_{\lambda}^L}(n)$ are referred to as core, core-valence and
valence correlation contributions, respectively.

\begin{center}
\begin{figure}[h]
\includegraphics[width=8.5cm,clip=true]{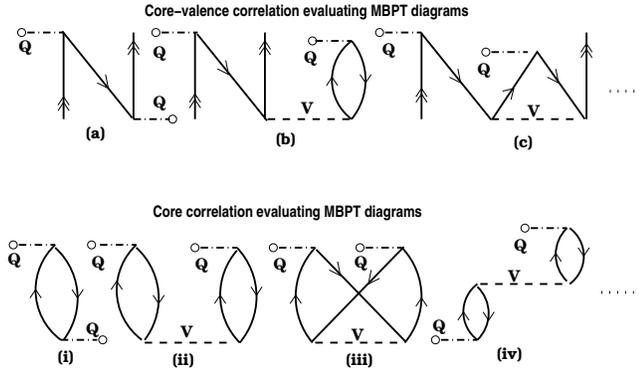}
\caption{Few important many-body perturbation theory (MBPT) diagrams used for the core-valence and core correlations estimation.} 
\label{fig1}
\end{figure}
\end{center}

In the sum-over-states approach as given in Eq. (\ref{eq25}), it is convenient 
to determine the low-lying singly excited states $|\Psi_m\rangle$ with respect
to the $|\Psi_n\rangle$ states that are of our interest for both the considered
ions. Contributions from these states corresponds to the above mentioned 
valence correlation contributions, which are the dominant ones compared to the
contributions to the part arising from higher level excited states that can be
estimated from a lower order perturbation theory. Contrasting to this 
contribution, it is not possible to estimate the core and core-valence
correlation contributions in a similar procedure. As it was stated above, 
methods describing in \cite{sahoo3, sahoo4, kallay} would be befitting 
to account them rigorously. Notwithstanding this fact, it will not alter 
the BBR shifts what is the primary intent of the work. This is because of
two reasons: First, the core correlation effect may be notable but in the BBR
shift estimation this contribution cancels out between the transition states.
However, we calculate them using the third order relativistic many-body
perturbation theory (MBPT(3)) to present the total polarizability results. 
Secondly, it is observed in the earlier
works that the core-valence contributions are minuscule in the dipole 
polarizability determinations \cite{BBR-arora-ca,BBR-arora-sr, sahoo3,
sahoo4} which is further verified below in the present work. So it is not
necessary to employ a powerful method at the cost of heavy computation to 
determine these tiny contributions for which we use again the MBPT(3) method to
estimate these contributions within the required precision level. In this approach,
we rewrite Eq. (\ref{eq25}) as
\begin{eqnarray}
\alpha_n^{Q_{\lambda}^L}  &=& \frac{1}{\alpha^{2(\lambda-1)}}\frac{2}{(2L+1)}\frac{1}{2J_n+1} \left < \Psi_n ||Q_{\lambda}^L || \Psi_n^{(1)} \right >,
\label{eq27}
\end{eqnarray}
where $|\Psi_n^{(1)}  >$ is the first order perturbed wave function for
$|\Psi_n\rangle$ due to $Q_{\lambda}^L$  which is obtained by solving the
following equation 
\begin{eqnarray}
(H- E_n) | \Psi_n^{(1)} > &=& (E_n^{(1)} - Q_{\lambda}^L) |\Psi_n >,
\end{eqnarray}
where $E_n^{(1)}$ is the energy correction due to $Q_{\lambda}^L$. Some of 
the important diagrams representing this equation for the core and
core-valence correlation effects evaluation are shown in Fig. \ref{fig1}.

The reduced transition matrix element of a physical operator $Q_{\lambda}^L$
between $|\Psi_f \rangle$ and $|\Psi_i \rangle$ in our approach is calculated
using the expression
\begin{eqnarray}
\frac{\langle \Psi_f || Q_{\lambda}^L || \Psi_i \rangle}{\sqrt{\langle \Psi_f|\Psi_f\rangle \langle \Psi_i|\Psi_i\rangle}} &=& \frac{\langle \Phi_f || \{ 1+ S_f^{\dagger}\} \overline{Q_{\lambda}^L } \{ 1+ S_i\} ||\Phi_i\rangle}{ \sqrt{{\cal N}_f {\cal N}_i}}, \nonumber \\
\label{eqn25}
\end{eqnarray}
where $\overline{ Q_{\lambda}^L}=e^{T^{\dagger}} Q_{\lambda}^L e^T$ and ${\cal N}_n = \langle \Phi_n | 
e^{T^{\dagger}} e^T + S_n^{\dagger} e^{T^{\dagger}} e^T S_n |\Phi_n\rangle $ involve two non-truncating 
series in the above expression. We have mentioned concisely the procedure for calculating these
expressions in the appendix in \cite{mukherjee}. We evaluate them keeping terms up to fourth order
in Coulomb interaction.

The single particle orbital reduced matrix elements for the corresponding 
M1 and E2 cases are taken as
\begin{eqnarray}
\langle \kappa_f\, ||\,m1\,||\, \kappa_i \rangle &=&  \frac {(\kappa_f+\kappa_i)} {\alpha}\langle - \kappa_f\, ||\,C^{(1)}\,||\,\kappa_i \rangle \nonumber \\ 
&& \int_0^{\infty} dr r (P_f(r)Q_i(r)+Q_f(r)P_i(r)), \ \ \ \ \ \ \
\end{eqnarray}
and 
\begin{eqnarray}
\langle \kappa_f\, ||\,e2\,||\,\kappa_i \rangle &=& \langle \kappa_f\, ||\,C^{(2)}\,||\, \kappa_i \rangle \nonumber \\
&& \int_0^{\infty} dr r^2  (P_f(r)P_i(r)+Q_f(r)Q_i(r)),\nonumber \\ 
\end{eqnarray}
where $P(r)$ and $Q(r)$ represent the radial parts of the
large and small components of the single particle Dirac orbitals,
respectively. The reduced Racah coefficients are given by
\begin{eqnarray}
\langle \kappa_f\, ||\, C^{(k)}\,||\, \kappa_i \rangle &=& (-1)^{j_f+1/2} \sqrt{(2j_f+1)(2j_i+1)} \ \ \ \ \ \ \ \ \nonumber \\
                  &&        \left ( \begin{matrix}
                              j_f & k & j_i \cr
                              1/2 & 0 & -1/2 \cr
                                       \end{matrix}
                            \right ) \pi(l_{\kappa_f},k,l_{\kappa_i}), \ \ \ \ \
\end{eqnarray}
with
\begin{eqnarray}
  \pi(l,m,l') &=&
  \left\{\begin{array}{ll}
      \displaystyle
      1 & \mbox{for } l+m+l'= \mbox{even}
         \\ [2ex]
      \displaystyle
        0 & \mbox{otherwise.}
    \end{array}\right.
\label{eqn12}
\end{eqnarray}

We have used Gaussian type orbitals (GTOs) to construct the single particle
orbitals for the Dirac-Fock ($|\Phi_0 \rangle$) wave function calculation.
The large and small components of the Dirac orbitals in this case are
expressed as
\begin{eqnarray}
P_{\kappa}(r) &=& \sum_k c_k^P r^{l_{\kappa}} e^{-\alpha_k r^2}
\end{eqnarray}
and
\begin{eqnarray}
Q_{\kappa}(r) &=& \sum_k c_k^Q r^{l_{\kappa}} \left ( \frac{d}{dr} + \frac{\kappa}{r} \right ) e^{-\alpha_k r^2},
\end{eqnarray}
where the summation over $k$ is for the total number of GTOs used in each symmetry,
$c_k^P$ and $c_k^Q$ are the normalization constants for the large and
small components, respectively, and we use $(\frac{d}{dr}+\frac{\kappa}{r})$
operator to expand the small component Dirac orbitals to maintain the kinetic
balance condition with its large component. In the present calculations, we
have considered 9 relativistic symmetries (up to $g$ symmetry) and 28 GTOs 
for each symmetry to generate the orbitals. In order to optimise
the exponents to describe orbitals from various symmetries in a smooth manner,
we use the even tempering condition
\begin{equation}
\alpha_k = \alpha_0 \beta^{k-1},
\end{equation}
where $\alpha_0$ and $\beta$ are two arbitrary parameters that are chosen
suitably for different symmetries. 

We have considered $\alpha_0=7.5\times10^{-4}$ for all the symmetries and 
$\beta$s are optimised to be $2.56$, $2.58$, $2.61$, $2.75$ and $2.83$ for $s$,
$p$, $d$, $f$ and $g$ orbitals, respectively. For the RCC calculations, we 
have considered excitations up to first $16s$, $16p$, $16d$, $14f$ and $13g$ 
orbitals while contributions from other virtual orbitals having large energies 
are neglected.

\section{Results and discussions}

\begin{table}
\caption{\label{ca1}Contributions to the $4s_{1/2}$ and $3d_{5/2}$ scalar ($\alpha^{M1}$) static polarizabilities in $^{43}$Ca$^+$ and
their uncertainties in units of $a_0^3$. The values of the corresponding M1 matrix elements for $\alpha Q_0^1$ are given in $ea_0$.}
\begin{ruledtabular}
\begin{tabular}{cccc}
\multicolumn{2}{c}{Transition} &
\multicolumn{1}{c}{Amplitude} &
\multicolumn{1}{c}{$\alpha^{M1}$}\\
\hline\\
\multicolumn{2}{c}{$\alpha_n$} & & \\
$4s_{1/2} \rightarrow$ & $5s_{1/2}$ & 0.0018(3) & $0.4(1) \times 10^{-5}$ \\
 & $6s_{1/2}$ & 0.0013(2) & $1.7(5) \times 10^{-6}$ \\
 & $3d_{3/2}$ & 0.0008(1)  & $3.1(8) \times 10^{-6}$    \\   
 & $4d_{3/2}$ & 0.0003(1)  & $0.9(7) \times 10^{-7}$   \\   
 & $5d_{3/2}$ & 0.0001(1)  & $0.8(20) \times 10^{-8}$   \\   
 & $6d_{3/2}$ & 0.0003(1)  & $0.6(5) \times 10^{-7}$   \\   
\multicolumn{2}{c}{$\alpha_c$}  &       &  $5.0(2) \times 10^{-5}$ \\
\multicolumn{2}{c}{$\alpha_{cn}$}  &       &  $-1.4(3) \times 10^{-8}$ \\
\multicolumn{2}{c}{$\alpha_{\rm{tail}}$}  &       & $1.0(2) \times 10^{-8}$   \\
\multicolumn{2}{c}{$\alpha_{\rm{total}}$} &       & $5.9(3) \times 10^{-5}$  \\
\hline\\
\multicolumn{2}{c}{$\alpha_n$} & & \\
$3d_{5/2} \rightarrow$ & $3d_{3/2}$ &  1.543(5)  & -957(6)   \\   
 & $4d_{3/2}$ &  0.004(1) & $8(4) \times 10^{-6}$   \\
 & $5d_{3/2}$ &  0.004(1) & $8(4) \times 10^{-6}$   \\
 & $6d_{3/2}$ &  0.003(1) & $2(2) \times 10^{-6}$   \\
 & $4d_{5/2}$ &  0.004(1) & $8(4) \times 10^{-6}$  \\
 & $5d_{5/2}$ &  0.009(2) & $3(1) \times 10^{-5}$   \\
 & $6d_{5/2}$ &  0.002(1) & $1(1) \times 10^{-6}$   \\
 & $5g_{7/2}$ & $3.2\times 10^{-7}$ & $\sim 0$ \\
 & $6g_{7/2}$ & $4.6\times 10^{-7}$ & $\sim 0$  \\
\multicolumn{2}{c}{$\alpha_c$}  &       & $5.0(2) \times 10^{-5}$  \\
\multicolumn{2}{c}{$\alpha_{cn}$}  &       &  $0.0$ \\
\multicolumn{2}{c}{$\alpha_{\rm{tail}}$}  &     & $-5.0(6) \times 10^{-6}$   \\
\multicolumn{2}{c}{$\alpha_{\rm{total}}$} &       & -957(6) \\
\end{tabular}   
\end{ruledtabular}
\end{table}
As has been emphasized before, precise estimation of the BBR shifts due to 
M1 and E2 transitions in Ca$^+$ and Sr$^+$ are the focus of this work. The 
uncertainties in these estimations are of two folds: First, the errors associated
with the considered excitation energies (EEs). Second, inaccuracies from the 
estimated transition matrix elements. To abate the uncertainties in the 
evaluation of the BBR shift for the atomic clock application, we use the
experimental EEs from the NIST database \cite{NIST1} for the important
singly excited states. Matrix elements for these states are mentioned 
explicitly along with their uncertainties and respective 
contributions to the the scalar polarizabilities in 
Tables~\ref{ca1}, ~\ref{sr1}, ~\ref{ca2}, and ~\ref{sr2}.
Contributions from the doubly excited states constructed within the 
configuration space spanned by the considered orbitals for our 
RCC calculations and from the higher singly excited states which 
are not mentioned specifically are accounted as $\alpha_{tail}$ using the
MBPT(3) method. As shown in Tables~\ref{ca1}, ~\ref{sr1}, ~\ref{ca2}, 
and ~\ref{sr2}, these contributions are rather diminutive 
to be concerned about. The uncertainties in these results and for the
transition matrix elements are estimated taking into account the neglected 
contributions and numerical calculations.

\begin{table}
\caption{\label{sr1}Contributions to the $5s_{1/2}$ and $4d_{5/2}$ scalar ($\alpha^{M1}$) static polarizabilities in $^{88}$Sr$^+$ and
their uncertainties in units of $a_0^3$. The values of the corresponding M1 matrix elements for $\alpha Q_0^1$ are given in $ea_0$.}
\begin{ruledtabular}
\begin{tabular}{cccc}
\multicolumn{2}{c}{Transition} &
\multicolumn{1}{c}{Amplitude} &
\multicolumn{1}{c}{$\alpha^{M1}$}\\
\hline\\ 
\multicolumn{2}{c}{$\alpha_n$} & & \\
$5s_{1/2} \rightarrow$ & $6s_{1/2}$ & $1.35(2) \times 10^{-3}$ & $ 2.79(8) \times 10^{-6}$\\
 & $7s_{1/2}$ & $1.24(1) \times 10^{-3}$ & $ 1.73(3) \times 10^{-6}$  \\
 & $8s_{1/2}$ & $1.00(1) \times 10^{-3}$ & $ 1.0(2) \times 10^{-6}$\\
 & $9s_{1/2}$ & $1.00(1) \times 10^{-3}$ & $ 9.4(2) \times 10^{-7}$  \\
 & $4d_{3/2}$ & $5.0(3) \times 10^{-5}$  & $ 1.3(2) \times 10^{-8}$   \\   
 & $5d_{3/2}$ & $1.0(1) \times 10^{-5}$ &  $ 1.4(3) \times 10^{-10}$  \\   
 & $6d_{3/2}$ & $1.0(1) \times 10^{-6}$ &  $ 1.1(2) \times 10^{-12}$  \\   
 & $7d_{3/2}$ & $3.0(2) \times 10^{-6}$ &  $ 8.8(6) \times 10^{-12}$  \\   
\multicolumn{2}{c}{$\alpha_c$}  &       & $1.0(1) \times 10^{-3}$  \\
\multicolumn{2}{c}{$\alpha_{cn}$}  &       &  $-2(1) \times 10^{-8}$  \\
\multicolumn{2}{c}{$\alpha_{\rm{tail}}$}  &       &  $1.0(1) \times 10^{-8}$  \\
\multicolumn{2}{c}{$\alpha_{\rm{total}}$} &       & $1.0(1) \times 10^{-3}$  \\
\hline \\
\multicolumn{2}{c}{$\alpha_n$} & & \\
$4d_{5/2} \rightarrow$ & $4d_{3/2}$ & 1.545(6) & $-208(2)$  \\   
 & $5d_{3/2}$ & 0.009(2) & $5(2) \times 10^{-5}$   \\
 & $6d_{3/2}$ & 0.006(1) & $1.5(5) \times 10^{-5}$   \\
 & $7d_{3/2}$ & 0.004(1) & $7(3) \times 10^{-6}$   \\
 & $5d_{5/2}$ & 0.003(1) & $6(4) \times 10^{-6}$  \\
 & $6d_{5/2}$ & 0.003(1) & $4(3) \times 10^{-6}$   \\
 & $7d_{5/2}$ & 0.002(1) & $2(2) \times 10^{-6}$   \\
 & $5g_{7/2}$ & $3.2(5)\times 10^{-7}$ & $\sim 0$  \\
 & $6g_{7/2}$ & $3.0(4) \times 10^{-7}$ & $\sim 0$    \\
\multicolumn{2}{c}{$\alpha_c$}  &       & $1.0(1) \times 10^{-3}$   \\
\multicolumn{2}{c}{$\alpha_{cn}$}  &       & $-2.5(2) \times 10^{-7}$   \\
\multicolumn{2}{c}{$\alpha_{\rm{tail}}$}  &     &  $-1.5(1) \times 10^{-7}$  \\
\multicolumn{2}{c}{$\alpha_{\rm{total}}$} &       & $-208(2)$ \\
\end{tabular}   
\end{ruledtabular}
\end{table}
We first present the static scalar $\alpha^{M1}$ polarizabilities 
in the Ca$^+$ and Sr$^+$ ions in Table \ref{ca1} and Table \ref{sr1}, 
respectively, for both the ground and $d_{5/2}$ states. As seen in these
tables, $\alpha^{M1}$s are very small for the corresponding ground states 
in both the ions, however they are reported in this paper for the completeness of the
results. Contributions from the core correlations are the largest in 
determining these results and they will be cancelled while estimating
the BBR shifts due to M1 multipole. Therefore, the ground state M1 
polarizability contributions in the considered ions can be neglected.
Essentially, the contributions to the $\alpha^{M1}$ polarizabilities
in the $d_{5/2}$ states are overwhelmingly dominant owing to the very 
small energy gap of their fine structure partner states
and contributions from all other states are small. The M1 matrix elements
between the $d_{3/2} - d_{5/2}$ transitions in the considered ions were
also reported earlier by us \cite{sahoo-lif} and they seem to be very
consistent. From our calculations, we obtain $\alpha^{M1}$ for the $3d_{5/2}$
and $4d_{5/2}$ states as $-957(6)$ a.u. and $-208(2)$  a.u. in the Ca$^+$ 
and Sr$^+$ ions, respectively.
The uncertainty in the final result is obtained by 
adding the uncertainty from each contribution in quadrature.
As shown in Table~\ref{sr1} the
uncertainty in the $d_{5/2}$ state polarizability value is dominated
by the uncertainty in the $d_{3/2} - d_{5/2}$ contribution.

\begin{table}
\caption{\label{ca2}Contributions to the $4s_{1/2}$ and $3d_{5/2}$ scalar ($\alpha^{E2}$) static polarizabilities in $^{43}$Ca$^+$ and
their uncertainties in units of $a_0^5$. The values of the corresponding E2 matrix elements for $Q_1^2$ are given in $ea_0^2$.}
\begin{ruledtabular}
\begin{tabular}{cccc}
\multicolumn{2}{c}{Transition} &
\multicolumn{1}{c}{Amplitude} &
\multicolumn{1}{c}{$\alpha^{E2}$}\\
\hline\\
\multicolumn{2}{c}{$\alpha_n$}  &       &    \\
$4s_{1/2} \rightarrow$ & $3d_{3/2}$ & 8.12(5)   & 212(3)  \\ 
 & $4d_{3/2}$ & 12.51(8)  & 121(2) \\ 
 & $5d_{3/2}$ & 3.89(4)   & 9.1(2) \\ 
 & $6d_{3/2}$ & 5.44(6)   & 16.2(4) \\   
 & $3d_{5/2}$ & 9.97(6)   &  318(3) \\
 & $4d_{5/2}$ & 15.30(9) & 181(2) \\
 & $5d_{5/2}$ & 4.75(5)   & 13.6(3) \\
 & $6d_{5/2}$ & 6.67(6)   & 24.2(4) \\
\multicolumn{2}{c}{$\alpha_c$}  &       &  6.15(8)  \\
\multicolumn{2}{c}{$\alpha_{cn}$}  &       &  0.0  \\
\multicolumn{2}{c}{$\alpha_{\rm{tail}}$}  &       & 5.36(5)   \\
\multicolumn{2}{c}{$\alpha_{\rm{total}}$} &       & 906(5) \\
 & & \\
\multicolumn{2}{c}{Other works} &       & 712.91$^a$, 871$^b$  \\
\hline\\
\multicolumn{2}{c}{$\alpha_n$}  &       &    \\
$3d_{5/2} \rightarrow$ & $4s_{1/2}$ & 9.97(6)  & -106(1) \\ 
& $5s_{1/2}$ & 4.99(5)  & 9.4(2) \\ 
& $6s_{1/2}$ & 1.22(2)  & 0.38(1) \\ 
& $3d_{3/2}$ & 3.90(4)  & -3657(75) \\   
& $4d_{3/2}$ & 4.32(5) &  6.3(1) \\
& $5d_{3/2}$ & 1.38(2) & 0.47(1) \\
& $6d_{3/2}$ & 1.08(1) & 0.25(5) \\
& $4d_{5/2}$ & 8.63(5) & 25.2(3) \\
& $5d_{5/2}$ & 2.75(3) &  1.87(4) \\
& $6d_{5/2}$ & 2.16(2) &  1.02(2) \\
 & $5g_{7/2}$ & 1.58(2)   & 0.56(1) \\
 & $6g_{7/2}$ & 1.04(1)  & 0.210(4) \\
 & $5g_{9/2}$ & 5.57(5)   & 7.0(1) \\
 & $6g_{9/2}$ & 3.67(4)  & 2.62(6) \\
\multicolumn{2}{c}{$\alpha_c$}  &       &  6.15(8)  \\
\multicolumn{2}{c}{$\alpha_{cn}$}  &       &  0.18(2)  \\
\multicolumn{2}{c}{$\alpha_{\rm{tail}}$}  &       & -4.29(7)   \\
\multicolumn{2}{c}{$\alpha_{\rm{total}}$} &       & -3706(75) \\
\end{tabular}   
\end{ruledtabular}
\begin{tabular}{ll}
References: & $^a$ \cite{sahoo-cpl}\\
  & $^b$ \cite{safronova}\\
\end{tabular}   
\end{table}
We now turn to the $\alpha^{E2}$ results. Our calculated results
for the ground and $3d_{5/2}$ states in Ca$^+$ are given in Table \ref{ca2}. 
Both the results are comparatively large with opposite signs. The largest
contribution to the ground state polarizability comes from the $3d$ states followed by
the $4d$ states. The core contribution is comparatively small and the tail
contribution to the ground state polarizability is zero due to the absence of occupied 
$d$ states in Ca$^+$. There are two more calculations available in 
the literature on the same using the RCC methods with different level of
approximations \cite{sahoo-cpl, safronova}. Result reported in 
\cite{sahoo-cpl} is {\it ab initio} and are obtained from a linear response
theory based calculation. Linear approximation in the RCC method
is being considered to evaluate the corresponding transition matrix elements
for the estimation of the ground state polarizability in \cite{safronova} using a 
sum-over-states approach like the present work. All the results are of same order
in magnitude. The {\it ab initio} result seems to be little lower than the 
results obtained from the sum-over-states approach due to the additional uncertainties
from the calculated energies that justifies for considering the experimental energies
in these calculations. However, it can be noticed that the differences in these results
will not alter the BBR shift results which can be apparent from the following finding
on the E2 polarizability contributions. To our knowledge, no other quadrupole
polarizability result is available for the $3d_{5/2}$ state in Ca$^+$ to compare with ours.
Again, contribution from its fine structure partner is the decisive factor for
the final result followed by a significant contribution from the ground state.
\begin{table}
\caption{\label{sr2}Contributions to the $5s_{1/2}$ and $4d_{5/2}$ scalar ($\alpha^{E2}_{0}$) static polarizabilities in $^{88}$Sr$^+$ and
their uncertainties in units of $a_0^5$. The values of the corresponding E2 matrix elements for $Q_1^2$ are given in $ea_0^2$.}
\begin{ruledtabular}
\begin{tabular}{cccc}
\multicolumn{2}{c}{Transition} &
\multicolumn{1}{c}{Amplitude} &
\multicolumn{1}{c}{$\alpha^{E2}$}\\
\hline\\
\multicolumn{2}{c}{$\alpha_n$}  &       &    \\
$5s_{1/2} \rightarrow$ & $4d_{3/2}$ & 11.25(7)   & 382(5)  \\ 
 & $5d_{3/2}$ & 12.87(8)  & 136(2) \\ 
 & $6d_{3/2}$ & 5.00(5)   & 16.2(3) \\ 
 & $7d_{3/2}$ & 3.11(4)   & 5.7(1) \\   
 & $4d_{5/2}$ & 13.91(8)   &  572(6) \\
 & $5d_{5/2}$ & 15.64(9) & 201(2) \\
 & $6d_{5/2}$ & 5.97(6)  & 23.3(4) \\
 & $7d_{5/2}$ & 3.76(4)   & 8.3(1) \\
\multicolumn{2}{c}{$\alpha_c$}  &       &  14.50(9)  \\
\multicolumn{2}{c}{$\alpha_{cn}$}  &       &  $-1.7(2) \times 10^{-8}$  \\
\multicolumn{2}{c}{$\alpha_{\rm{tail}}$}  &       & 6.35(8)   \\
\multicolumn{2}{c}{$\alpha_{\rm{total}}$} &       & 1366(9) \\
\hline\\
\multicolumn{2}{c}{$\alpha_n$}  &       &    \\
$4d_{5/2} \rightarrow$ & $5s_{1/2}$ & 13.91(8)  & -191(2) \\ 
 & $6s_{1/2}$ & 1.31(1)  & 0.77(1) \\ 
 & $7s_{1/2}$ & 1.90(2)  & 1.05(2) \\ 
 & $8s_{1/2}$ & 0.96(1) & 0.231(5)\\
 & $9s_{1/2}$ & 0.72(1)  & 0.119(4) \\
 & $4d_{3/2}$ & 6.08(6)  & -1930(38) \\   
 & $5d_{3/2}$ & 5.63(6) &  12.1(3) \\
 & $6d_{3/2}$ & 1.83(2) & 0.92(2) \\
 & $7d_{3/2}$ & 1.14(1) & 0.315(6) \\
 & $5d_{5/2}$ & 11.17(7) & 47.4(6) \\
 & $6d_{5/2}$ & 3.65(5) &  3.7(1)\\
 & $7d_{5/2}$ & 2.28(3) &  1.27(3) \\
 & $5g_{7/2}$ & 2.73(3)  & 1.76(4) \\
 & $6g_{7/2}$ & 2.21(2)  & 1.11(2) \\
 & $5g_{9/2}$ & 9.64(7)   & 22.0(3) \\
 & $6g_{9/2}$ & 7.80(6)  & 13.7(2) \\
\multicolumn{2}{c}{$\alpha_c$}  &       & 14.50(9)   \\
\multicolumn{2}{c}{$\alpha_{cn}$}  &       &  $0.24(3)$  \\
\multicolumn{2}{c}{$\alpha_{\rm{tail}}$}  &       & $-4.83(5)$   \\
\multicolumn{2}{c}{$\alpha_{\rm{total}}$} &       & -2005(38) \\
\end{tabular}   
\end{ruledtabular}
\end{table}
In Table \ref{sr2}, we present $\alpha^{E2}$ results for the ground and 
$4d_{5/2}$ states in Sr$^+$. The magnitudes of the ground state result
in this ion is larger than Ca$^+$, while for the corresponding $d_{5/2}$
it is other way around.  There are no results
available to the best of our knowledge to compare our results with them.
The trend of the contributions from different transitions is almost similar
for corresponding states in both the ions. 
We add the uncertainty from each contribution in quadrature to obtain the final uncertainty in E2 polarizability values.

Using the above values of the polarizabilities, we obtain the BBR shift
due to the M1 multipole for the $4s \ ^2S_{1/2} \rightarrow 3d \ ^2D_{5/2}$
transition in Ca$^{+}$ at the room temperature (300 K) to be $4.38(3)
 \times 10^{-4}$ Hz. Similarly, this result comes out to be $9.50(7) 
\times 10^{-5}$ Hz for the $5s \ ^2S_{1/2} \rightarrow 4d \ ^2D_{5/2}$
transition in Sr$^{+}$. Contributions from the E2 multipole are very small 
and below the uncertainties of the above results and hence can be neglected 
for the present purpose of the work.  However, the reported quadrupole
polarizabilities for all the considered states may be useful elsewhere.
It has to be noticed that the BBR shifts due to the E1 multipole are 
0.38(1) Hz \cite{sahoo4,safronova} and $0.22(1)$ Hz \cite{sahoo3} for the
corresponding transitions in Ca$^{+}$ and Sr$^{+}$, respectively.

\section{Conclusion}

In summary, we have estimated the black-body radiation shifts due to
the magnetic dipole and electric quadrupole multipoles for the 
$4s \ ^2S_{1/2} \rightarrow 3d \ ^2D_{5/2}$ and $5s \ ^2S_{1/2} \rightarrow 
4d \ ^2D_{5/2}$ transitions in the singly ionized calcium and strontium,
respectively. The contribution due to the former is the decisive in this
case. Nevertheless, the reported polarizabilities for the considered states
which are rarely studied in the above ions may also be useful for other purposes.
It may be imperative to contemplate the reported shifts which are given as 
$4.38(3) \times 10^{-4}$ Hz and $9.50(7) \times 10^{-5}$ Hz in
the considered ions to achieve the $10^{-18}$ precision uncertainty in the
proposed clock experiments.

\section*{Acknowledgement}
The work of B.A. was supported by the Department of Science and
Technology, India. Computations were carried out using 3TFLOP 
HPC Cluster at Physical Research Laboratory, Ahmedabad.

\appendix

\section{Farley \& Wing's functions}

With the aid $|y| >>1$, the following the expression
\begin{eqnarray}
F_L(y)&=& \frac{1}{\pi}\frac{L+1}{L(2L+1)!!(2L-1)!!} \nonumber\\
&\times &\int_0^{\infty}\left( \frac{1}{y+x}+\frac{1}{y-x}\right)\frac{x^{2L+1}}{e^x-1} dx,
\end{eqnarray}
gives for $L=1$ as
\begin{eqnarray}
	F_1(y)&=& \frac{2}{3\pi}\int_0^{\infty}\left(\frac{1}{y+x}+\frac{1}{y-x}\right)\frac{x^3}{e^x-1}dx \nonumber \\
		  &=&\frac{2}{3\pi}\left(\frac{2}{y}\int_0^{\infty}\frac{x^3}{e^x-1}dx +\frac{2}{y^3}\int_0^{\infty}\frac{x^5}{e^x-1}dx\right) . \nonumber \\
\end{eqnarray}

Further by using the definite integral value 
\begin{equation}
\int_0^{\infty}\frac{x^{2n-1}}{e^{px}-1} dx = (-1)^{n-1}\left(\frac{2\pi}{p}\right)^{2n}\frac{B_{2n}}{4n},
\label{eqn10}
\end{equation}
where $B_{2n}$ is the Bernoulli number, $F_1(y)$ reduces to
\begin{equation}
	F_1(y)=\frac{4\pi^4}{45y}.
\label{eqnf}
\end{equation}

Similarly for $L=2$, the above expression turns out to be
\begin{eqnarray}
F_2(y)&=&\frac{1}{30\pi}\int_0^{\infty}\left(\frac{1}{y+x}+\frac{1}{y-x}\right)\frac{x^5}{e^x-1}dx\nonumber \\
        &=&\frac{1}{15y}\int_0^{\infty}\frac{x^5}{e^x-1}dx +\frac{2}{y^3}\int_0^{\infty}\frac{x^7}{e^x-1}dx\nonumber \\
&=&\frac{8\pi^5}{945y}. 
\label{eqn16}
\end{eqnarray}

\section{Square of the matrix element}

In our approach, we write
\begin{eqnarray}
|\Psi_n \rangle &=& a_n^{\dagger} \Omega_c |\Phi_0 \rangle + \Omega_{cn} |\Phi_v\rangle + \Omega_n |\Phi_v \rangle.
\end{eqnarray}

With this expression, the square of the matrix element of any arbitrary 
operator $O$ can be expressed as
\begin{eqnarray}
\left<\Psi_n|O|\Psi_m\right>^2 &=& \left<\Psi_n|O|\Psi_m\right>\left<\Psi_m|O|\Psi_n\right> \nonumber \\
 &=& \left<\Phi_0| \Omega_c^{\dagger} O \Omega_m \Omega_m^{\dagger} O  \Omega_c |\Phi_0\right>  \nonumber \\ 
 && + \left<\Phi_0| \Omega_c^{\dagger} O \Omega_{cm} \Omega_{cm}^{\dagger} O  \Omega_c |\Phi_0\right>  \nonumber \\ 
 && + \left<\Phi_n| \Omega_{cn}^{\dagger} O \Omega_c \Omega_c^{\dagger} O  \Omega_{cn} |\Phi_n\right> \nonumber \\ 
 && + \left<\Phi_n| \Omega_n^{\dagger} O \Omega_c \Omega_c^{\dagger} O  \Omega_n |\Phi_n\right> \nonumber \\ 
 && + \left<\Phi_n| \Omega_n^{\dagger} O \Omega_{cm} \Omega_{cm}^{\dagger} O  \Omega_n |\Phi_n\right>  \nonumber \\ 
 && + \left<\Phi_n| \Omega_n^{\dagger} O \Omega_m \Omega_m^{\dagger} O  \Omega_n |\Phi_n\right>,
\label{mateq}
\end{eqnarray}
where we have facilitated the generalized Wick's theorem to derive these 
terms and assumed all the operators are in normal order form and only the
connected terms are survived. For simplicity the first two terms, the second 
term and the last three terms are categorized into core (c), core-valence (cn) 
and valence (n) contributions, respectively; i.e. in an abbreviate form it 
is given as
\begin{eqnarray}
\left<\Psi_n|O|\Psi_m\right>^2 &=& \left<\Psi_n|O|\Psi_m\right>_c^2 + \left<\Psi_n|O|\Psi_m\right>_{cn}^2 \nonumber \\ && + \left<\Psi_n|O|\Psi_m\right>_n^2.
\end{eqnarray}


\end{document}